\documentclass[aps, prb, twocolumn, superscriptaddress, english]{revtex4-2}

\usepackage{graphicx}
\usepackage{cancel}
\usepackage{enumerate}
\usepackage{hyperref}
\usepackage{textcomp}
\usepackage{amsmath}
\usepackage{amssymb}
\usepackage{pgf}
\usepackage{soul}
\usepackage{subfigure}
\usepackage[english]{babel}
\usepackage[normalem]{ulem}

\def\ket#1{| #1\rangle}

\newcommand{\nep}{\textrm{e}}

\newcommand{\opf}[1]{{\hat{f}^{\phantom \dagger}}_{#1}}
\newcommand{\opfdag}[1]{{\hat{f}^{\dagger}}_{#1}}
\newcommand{\opc}[1]{{\hat{c}^{\phantom \dagger}}_{#1}}
\newcommand{\opcdag}[1]{{\hat{c}^{\dagger}}_{#1}}

\newcommand{\opd}[1]{{\hat{d}^{\phantom \dagger}}_{#1}}
\newcommand{\opddag}[1]{{\hat{d}^{\dagger}}_{#1}}

\newcommand{\opgamma}[1]{{\hat{\gamma}^{\phantom \dagger}}_{#1}}

\newcommand{\Ham}{\widehat{H}}

\newcommand{\Up}{{\uparrow}}
\newcommand{\Dn}{{\downarrow}}

\graphicspath{{./}{./Figure/}}

\begin{document}
\title{The topological Kondo model out of equilibrium}
\author{Matteo M. Wauters}
\affiliation{Niels Bohr International Academy and Center for Quantum Devices, Niels Bohr Institute, Copenhagen University, Universitetsparken 5, 2100 Copenhagen, Denmark}
\affiliation{ Pitaevskii BEC Center, Department of Physics, University of Trento, Via Sommarive 14, I-38123 Povo, Trento, Italy }

\author{Chia-Min Chung}
\affiliation{Department of Physics, National Sun Yat-sen University, Kaohsiung 80424, Taiwan}
\affiliation{Center for Theoretical and Computational Physics, National Sun Yat-Sen University, Kaohsiung 80424, Taiwan}
\affiliation{Physics Division, National Center for Theoretical Sciences, Taipei 10617, Taiwan}

\author{Lorenzo Maffi}
\affiliation{Niels Bohr International Academy and Center for Quantum Devices, Niels Bohr Institute, Copenhagen University, Universitetsparken 5, 2100 Copenhagen, Denmark}

\author{Michele Burrello}
\affiliation{Niels Bohr International Academy and Center for Quantum Devices, Niels Bohr Institute, Copenhagen University, Universitetsparken 5, 2100 Copenhagen, Denmark}

\begin{abstract}
The topological Kondo effect is a genuine manifestation of the nonlocality of Majorana modes. We investigate its out-of-equilibrium signatures in a model with a Cooper-pair box hosting four of these topological modes, each connected to a metallic lead. 
Through an advanced matrix-product-state approach tailored to study the dynamics of superconductors, we simulate the relaxation of the Majorana magnetization, which allows us to determine the related Kondo temperature, and we analyze the onset of electric transport after a quantum quench of a lead voltage.
Our results apply to Majorana Cooper-pair boxes fabricated in double nanowire devices and provide nonperturbative evidence of the crossover from weak-coupling states to the strongly correlated topological Kondo regime. The latter dominates at the superconductor charge degeneracy points and displays the expected universal fractional zero-bias conductance. 

\end{abstract}

\maketitle
The engineering of Majorana zero-energy modes (MZMs) in hybrid superconducting-semiconducting devices has been the core of strenuous theoretical and experimental activities for the last two decades \cite{Prada2020,Beenakker2020,Flensberg2021}.
The detection of these subgap modes relies primarily on tunneling spectroscopy applied to a rich variety of platforms, which, however, cannot provide direct evidence of the most intriguing properties of MZMs, namely their nonlocal and anyonic features. 
Hence, it is desirable to devise a new generation of experiments that balances the constraints imposed by the current technological limitations and the pursuit of MZM evidence beyond spectroscopy.

\begin{figure}
    \centering
    \includegraphics[width=6cm]{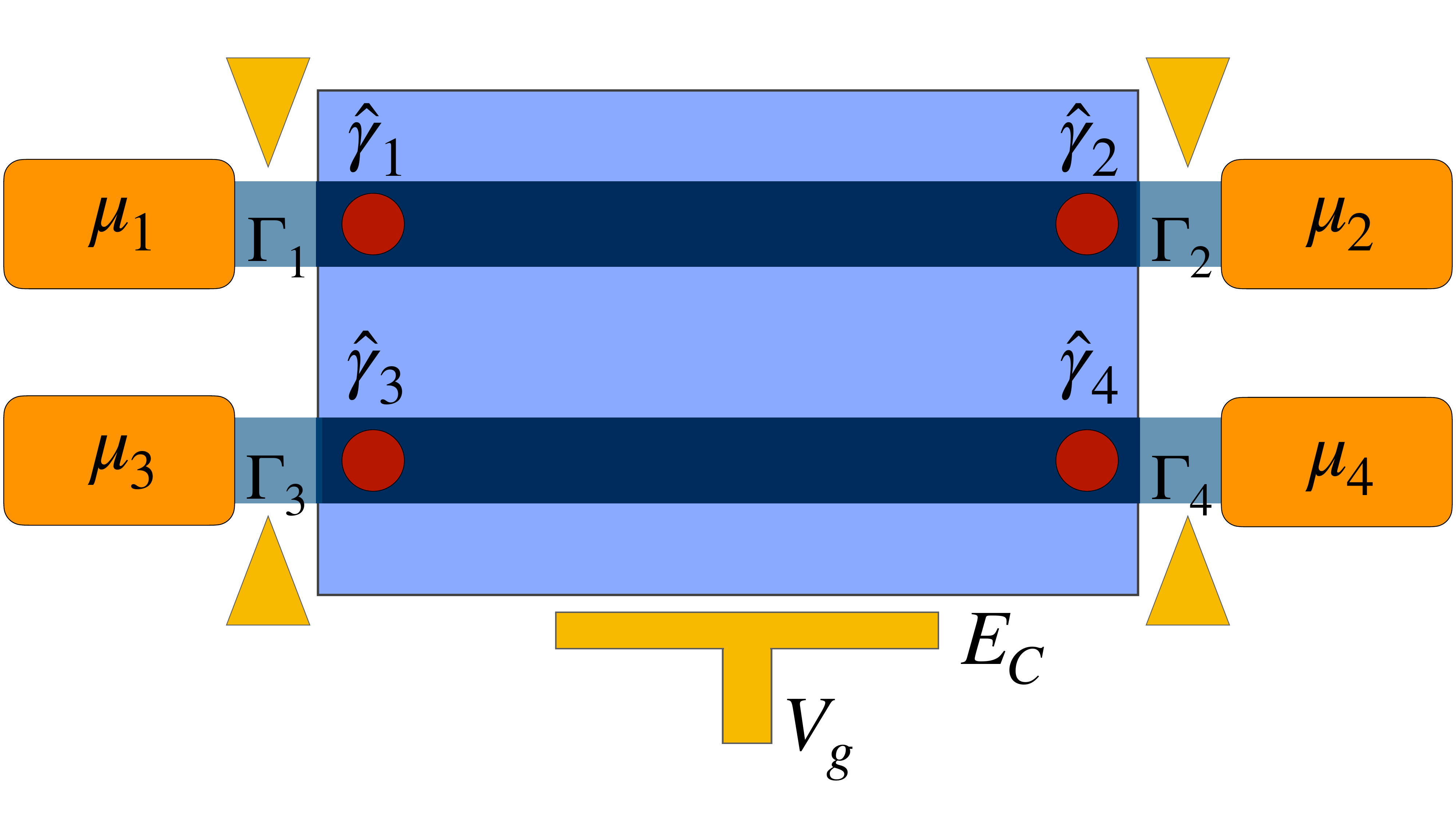}
    \caption{Schematics of the system: two p-wave superconducting nanowires with MZMs at the edges are coupled by a SC island  (blue) with charging energy $E_c$. Voltage gates (yellow) tune the island induced charge, $n_g \propto V_g$,  and the coupling rates $\Gamma_\alpha$ with the leads (orange). Each MZM is coupled with a single normal lead at chemical potential $\mu_\alpha$.}
    \label{fig:TKE_sketch}
\end{figure}

In this respect, the topological Kondo effect (TKE) \cite{Beri2012,Beri2013,Altland2013} plays a crucial role: on one side, it is a transport signature of MZMs well-suited for experimental observations; on the other, it results from their nonlocality and can hardly be confused with phenomena originating by nontopological subgap states \cite{Liu2021}.
The TKE is predicted to emerge in multiterminal devices where $M$ external leads are coupled to a Majorana Cooper-pair box hosting four MZMs and characterized by a charging energy $E_c$ (Fig. \ref{fig:TKE_sketch}). 
The TKE manifests itself as a universal nonlocal zero-bias conductance $dI_\alpha/dV_{\beta\neq \alpha}$ quantized at values $2e^2/Mh$. 
Such conductance is approached at low temperatures in correspondence of both the Coulomb valleys and peaks of the related devices~\cite{Michaeli2017}, as derived from renormalization group (RG) analyses of effective low-energy models describing the Majorana Cooper-pair box coupled to $M$ leads \cite{Beri2012,Beri2013,Altland2013,Altland2014,Galpin2014,Mora2016,Michaeli2017,Buccheri2022}.

In this letter, we address the experimental observability of the TKE from a more elementary and microscopic perspective and we examine its onset out of equilibrium. We characterize the quantum-quench dynamics of a minimal interacting fermionic model that includes both MZMs and quasiparticle excitations above the superconducting gap.
The time evolution is determined by electrons tunneling from the leads to the superconducting (SC) island, and, differently from previous TKE studies \cite{Beri2012,Altland2013,Galpin2014,Altland2014b,Buccheri2015,Beri2017,Papaj2019}, we apply matrix-product-state (MPS) simulations \cite{Chung_PRB2022} which do not rely on any perturbative approximation of this coupling nor on a clear energy scale separation. 
This technique allows us to examine the crossover between the predicted weak-coupling and topological Kondo strong-coupling regimes and estimate the related topological Kondo temperature $T_K$.

The model we propose provides a minimal description of Majorana Cooper-pair boxes engineered from nanowires. 
Recent developments in the fabrication of parallel InAs nanowires hybridized with Al \cite{Kanne2022,Vekris_NanoLett2022} make these platforms suitable to combine all the necessary elements to implement of the topological Kondo model.
It is therefore important to investigate its transport signatures as a function of the most relevant experimental parameters: lead voltages, charge induced on the SC island, and lead-island couplings.

\paragraph{Model and methods.-} 
We consider a TKE model composed by two parallel 1D topological superconductors coupled by a common floating SC island with charging energy $E_c$ and charge $n_g$ induced by the potential $V_g$ (Fig. \ref{fig:TKE_sketch}). 
It describes two nanowires with strong spin-orbit coupling subject to a proximity-induced SC pairing and a suitable Zeeman interaction, which provide a common route to engineer MZMs~\cite{Oreg2010,Lutchyn2010}. 
Their low-energy physics is captured by spinless fermions subject to an emergent p-wave SC pairing $\Delta_P$. As a result, four MZMs $\lbrace \gamma_\alpha\rbrace_{\alpha=1,\dots,4}$ form at the nanowire edges, each coupled to a spinless normal lead. 
The tunneling rates $\Gamma_\alpha$ between leads and MZMs can be switched off to change the number of terminals $M \le 4$ coupled to the system.

The simplest description for each SC nanowire is a zero-bandwidth model~\cite{Grove-Rasmussen2018,Vaitiekenas_PRB2022,Souto_PRB2022}, where the lowest energy level is the subgap state defined by two Majorana operators, while the higher energy state represents Bogoliubov quasiparticles above the SC gap. 
This corresponds to a 2-site Kitaev chain for each nanowire, with each of the four constituting fermionic sites tunnel-coupled to one of the leads.
This system defines the SC box sketched in Fig.~\ref{fig:TKE_sketch}, with Hamiltonian $\Ham = \Ham_{\rm sys}+\Ham_L+\Ham_{\rm t}$. $\Ham_{\rm sys}$ describes the Majorana Cooper-pair box \cite{vanHeck2012,Plugge2017}:
\begin{align} \label{Hsyst}
\Ham_{\rm sys}=\sum_{\sigma,n} \epsilon_{n,\sigma} \opfdag{n,\sigma}\opf{n,\sigma}  + E_c(\hat{N}-n_g)^2 \ , 
\end{align}
where $\sigma=\Up,\Dn$ labels the upper and lower nanowires and $n=0,1$ the two quasiparticle states in each of them \cite{Supplemental}.
$\hat{N}$ is the total box charge with respect to an arbitrary offset. It includes both its Cooper pairs and the electrons in the nanowires.

The two zero-energy quasiparticles are generated by the combinations of MZMs $\opf{0,\uparrow}  = (\opgamma{1} - i \opgamma{2})/2$ and $\opf{0,\downarrow} = (\opgamma{3} - i \opgamma{4})/2$. We label the four corresponding low-energy states by  $|n_\uparrow n_\downarrow \rangle$, with $\hat{n}_\sigma = \opfdag{0,\sigma}\opf{0,\sigma}$.
The charging energy splits them into two degenerate pairs with different fermionic parity $(-1)^{\hat{N}}$.
Hereafter, we set equal SC pairing and nanowire hopping, $\Delta_P=t_{\rm sys}=0.5 t_0$, and introduce a small potential $\mu_{\rm sys}=0.01t_0$ \cite{Supplemental}.

The leads are modeled by Wilson chains \cite{Wilson_RevModPhys75,DaSilva_PRB2008,Chung_PRB2022}
\begin{equation} \label{Hleads}
\Ham_{L} = \sum^4_{\alpha=1} \sum_{l=1}^\mathcal{L} \left[ -t_0 \nep^{-(l-1)/\xi} \opcdag{\alpha,l+1}\opc{\alpha,l} +{\rm h.c.} \right] - \mu_\alpha \opcdag{\alpha,l}\opc{\alpha,l} \ , 
\end{equation}
with $t_0$ being the \textit{bare} hopping amplitude which sets their bandwidth and constitutes the largest energy scale. The hopping decay length $\xi$ is an auxiliary variable allowing us to tune the resolution at small energies by modifying the lead level spacing~\cite{DaSilva_PRB2008,Chung_PRB2022,Wauters_arxiv2023}. The chemical potentials $\mu_\alpha$ are used to bring the system out of equilibrium and study transport properties.

The tunneling Hamiltonian between the leads and the system is
\begin{equation}\label{eq:tunn}
\Ham_{\rm t} = -\sum^4_{\alpha=1}\sum_{\sigma,n} J_{\alpha} \left[ \left( u_{\alpha,\sigma,n} \opfdag{\sigma,n}+v_{\alpha,\sigma,n} \opf{\sigma.n}\right)\opc{\alpha,1}  +{\rm H.c.} \right],
\end{equation}
where $u_{\alpha,\sigma,n}$ ($v_{\alpha,\sigma,n}$) is the particle (hole) projection of $\opf{\sigma,n}$ on the real-space site coupled to lead $\alpha$.
The amplitudes $J_\alpha$ determine the effective tunneling rates $\Gamma_\alpha=\frac{J_\alpha^2}{2t_0}$.

In our simulations, we map the system into an MPS by following Refs. \cite{Chung_PRB2022,Wauters_arxiv2023}.
Each MPS site represents a single-particle eigenstate of either the leads or the nanowires (Bogoliubov quasiparticles for nanowires) and we order them based on their energy.
The charge $\hat{N}$ is encoded in an auxiliary site \cite{keselman2019,Supplemental}.
The real-time dynamics is simulated using the time-dependent variational principle algorithm~\cite{Haegeman_PRL2011,Haegeman2016,Yang2020} from the ITensor library~\cite{itensor,source}.
We set the MPS truncation error $\sim 5 \cdot 10^{-8}$, corresponding to a maximum bond dimension $\chi\lesssim 2000$.

\paragraph{Relaxation towards equilibrium.-} 
\begin{figure}[t]
    \centering
    \includegraphics[width=0.75\columnwidth]{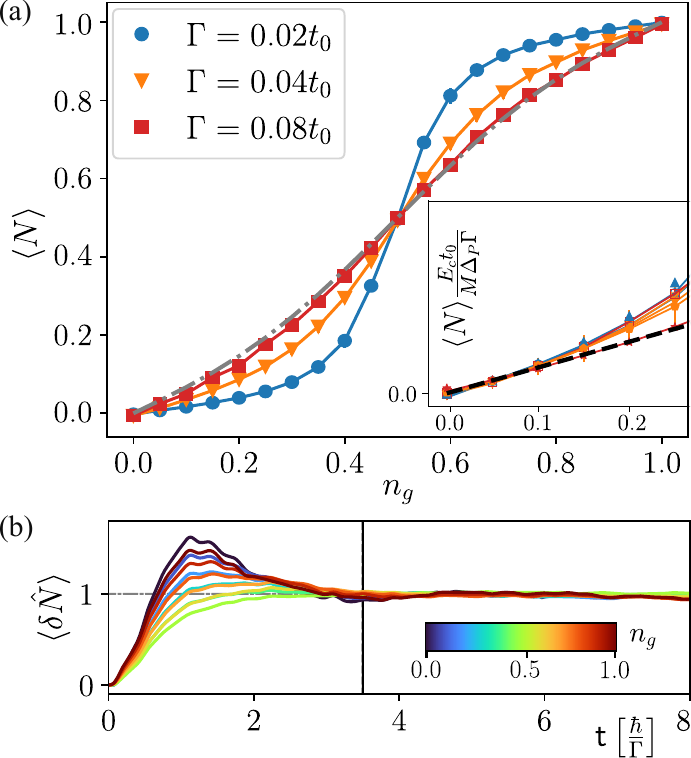}
    \caption{(a) Equilibrium charge versus $n_g$, for $M=3$, $E_c=0.2t_0$. The gray dot-dashed line corresponds to Eq.~\eqref{eq:Nvs_ng_sc} for $\Gamma=0.08t_0$.
    Inset: data for different values of $E_c$ ($0.2t_0$ and $0.4t_0$) and $M=3,\ 4$ in the weak-coupling regime, rescaled by $\frac{M\Delta_P\Gamma}{E_c t_0}$. The dashed black line corresponds to  $\langle \hat{N} \rangle= \frac{M\Delta_P\Gamma}{E_c t_0}n_g$.
    (b) Charge relaxation for different values of $n_g\in[0,1]$, $E_c=0.2t_0$ and $\Gamma=0.04t_0$. 
    All data are obtained with $\mathcal{L}=64$ and $\xi=16$.
    }
    \label{fig:charge}
\end{figure}
In the Kondo problem dynamics, the formation of strong correlations and the Kondo screening cloud occurs over a timescale given by $T_K^{-1}$ \cite{Lobaskin2005,Anders2005,Wauters_arxiv2023,Cavalcante_PRB2023}. 
Therefore, the relaxation after a quantum quench offers a useful probe to estimate the Kondo temperature and detect strongly correlated states.

In the following, we consider the dynamics of the system in a post-quench non-equilibrium quasi-steady state (NEQSS). In this regime, the time-evolution fulfills a Lieb-Robinson bound \cite{liebrobinson,bonnes2014} such that the dynamics is not affected by finite-size limitations until the signal propagates to the edge of the simulated system. Due to this, the information acquired from the analysis of the NEQSS well represents the out-of-equilibrium physics in the thermodynamic system (see, for instance, Refs. \cite{bertini2016,Viti_2016,Essler_2016,Ljubotina_SciPost2019}), as verified also by applying our protocol to superconducting interacting scatterers \cite{Chung_PRB2022} and the Anderson impurity model \cite{Wauters_arxiv2023}.

The first quench protocol we consider aims at observing the relaxation of the Majorana Cooper-pair box caused by $\Ham_{\rm t}$. 
The box is initialized in the ground state $|00\rangle$ ($N=0$ for $n_g < 0.5)$ or $|10\rangle$ $(N=1$ for $n_g>0.5)$ and is decoupled from the leads, which are set at half-filling. 
At time ${\sf t}=0$, the couplings $\Gamma$ are suddenly turned on and the device begins relaxing toward equilibrium. 
To characterize this relaxation, we analyze the average island charge $\langle \hat{N}(t)\rangle$,  and the effective Majorana magnetization \cite{Beri2012,Altland2014b} $\langle \hat{Z}_{\rm eff} (t)\rangle \equiv \langle i\opgamma{3}\opgamma{4}(t)\rangle= 1 -2 \langle \hat{n}_\downarrow(t)\rangle $.

The observed dependence of $\langle \hat{N} \rangle$ on $n_g$ after equilibration (Fig. \ref{fig:charge}) shows the crossover between the weak-coupling and the strong-coupling regime.
Following Ref.~\cite{Lutchyn2020}, we characterize the weak-coupling regime at $n_g\sim 0$ by the slope of $\langle \hat{N} \rangle = \frac{M\Delta_P\Gamma}{E_c t_0}n_g$:
For weak $\Gamma$, the charge datasets corresponding to different choices of $E_c$ and $M$ \footnote{The simulations with $M=3$ are performed by setting $J_2=0$.} exhibit a good agreement with the expected linear dependence [inset of Fig.~\ref{fig:charge}(a)]. On the other hand, the sinusoidal correction derived for the strong-coupling regime~\cite{Lutchyn2020},
\begin{equation}\label{eq:Nvs_ng_sc}
    \langle \hat{N}\rangle = n_g -\left( \frac{E_c}{\Delta_P}\sqrt{1-\Gamma/t_0}\right)^M \sin(2\pi n_g) ,
\end{equation}
closely matches the numerical data for the highest value of the tunneling rate $\Gamma=0.08t_0$ [gray dot-dashed line and red squares in Fig.~\ref{fig:charge}(a)], thus suggesting the emergence of Kondo correlations.

Importantly, the relaxation timescale of $\langle\hat{N}\rangle$ depends on the ratio $\Gamma/E_c$ but not on the induced charge $n_g$, as shown in Fig.~\ref{fig:charge}(b) which displays the time dependence of the relative charge variation, 
\begin{equation}
    \langle \delta \hat{N} (t)\rangle =  \frac{|\langle \hat{N} (t) \rangle-\langle \hat{N} (0) \rangle|}{|\langle \hat{N} (t\to \infty) \rangle-\langle \hat{N} (0) \rangle|} \ .
\end{equation}
The vertical line marks the equilibration time and different curves, corresponding to different values of $n_g \in~[0,1]$, converge to $\langle \delta \hat{N} (t)\rangle=1$ on similar timescales.

\begin{figure}
    \centering
    \includegraphics[width=\columnwidth]{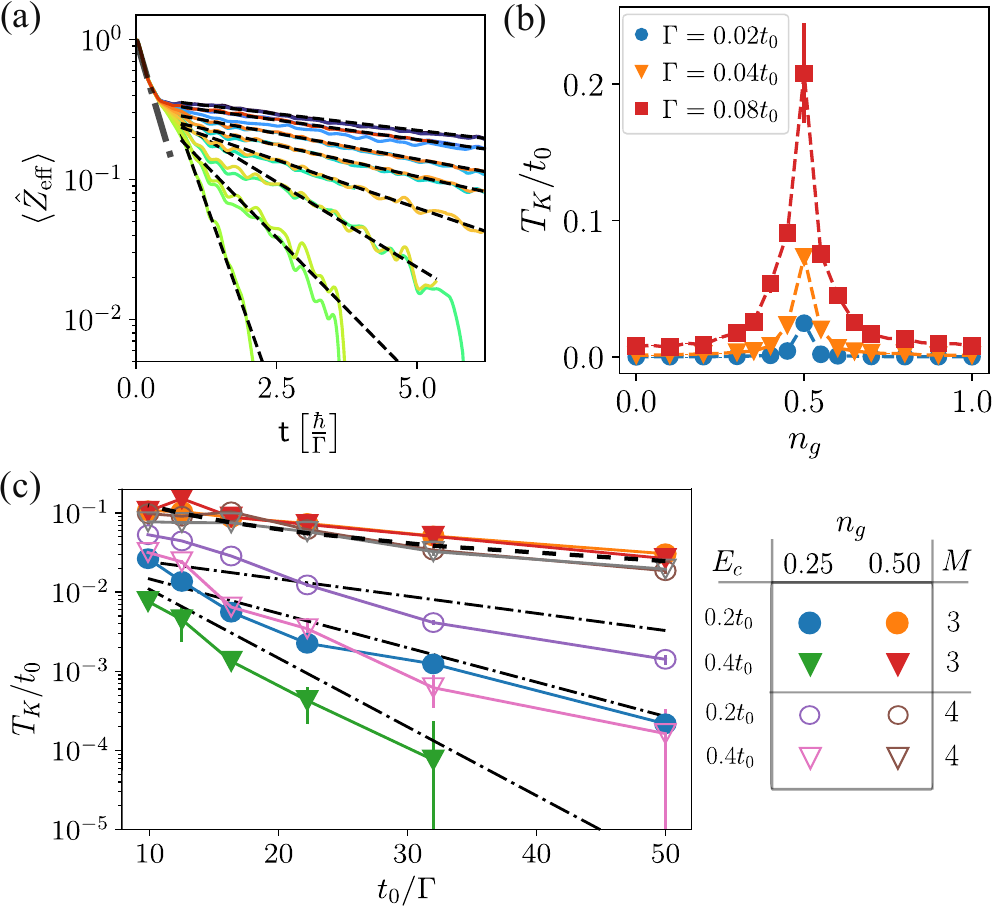}
    \caption{(a): Dynamics of the Majorana magnetization for different values of $n_g \in [0,1]$; $E_c=0.2t_0$ and $\Gamma=0.08t_0$. 
    The dot-dashed line marks the decay rate $\Gamma$.
    (b) $T_K$ extracted as the relaxation rate of $\langle \hat{Z}_{\rm eff}\rangle$ ---dashed lines in panel (a)--- as a function of $n_g$.
    (c) $T_K$ as a function of $t_0/\Gamma$ at $n_g=0.5$ and in the even-parity Coulomb valley ($n_g=0.25$). Dot-dashed lines correspond to Eq.~\eqref{eq:TK}, whereas the dashed line marks the degeneracy point scaling $T_K \sim M\Gamma$. A prefactor $C\sim 0.2$ has been manually set to approximately match the data.
    All data are obtained with $\mathcal{L}=64$, $\xi=16$.}
    \label{fig:Magnetization}
\end{figure}

The magnetization, instead, displays a remarkably different behavior [Fig.\ref{fig:Magnetization}(a)]. At short times, ${\sf t}<{\hbar}/{\Gamma}$, the relaxation is dominated by the fast rate $\Gamma$ (dot-dashed line) independently of both $E_c$ and $n_g$. 
Then, a second timescale emerges, which depends on both $\Gamma$ and the energy difference $\delta E(n_g)=E_c|1-2n_g|$ between the charge sectors $N=0$ and $N=1$. 
The black dashed lines in Fig.~\ref{fig:Magnetization}(a) are exponential fits of these slower decays.
This behavior is analogous to the magnetization relaxation in the Anderson impurity model~\cite{Anders2005,He_PRB2017,Wauters_arxiv2023}, suggesting that this longer timescale is associated with the energy scale $T_K$ of the emerging TKE. 

The comparison of Figs. \ref{fig:Magnetization}(a) and \ref{fig:charge}(b) underlines that this Kondo timescale characterizes only the Majorana magnetization but not the charge; $\langle \hat{Z}_{\rm eff}\rangle$ constitutes indeed one of the effective Pauli operators at the heart of the definition of the TKE, whereas $\langle\hat{N}\rangle$ depends only on the fermionic parity of the SC island. 
Therefore, we interpret this charge - ``spin" separation as evidence of the TKE emergence.

Next, we analyze the dependence of the so-derived $T_K$ on $n_g$, $\Gamma$, and $E_c$.
Figure~\ref{fig:Magnetization}(b) depicts the fitted $T_K$ as a function of $n_g$ for different values of $\Gamma$ and $E_c=0.2t_0$. 
As expected from RG analyses, $T_K$ is larger at the charge degeneracy point where it is proportional to $M\Gamma$, consistently with Ref. \cite{Lutchyn2020}.
In the Coulomb valleys, instead,  $T_K$ is qualitatively compatible with standard RG predictions \cite{Lutchyn2020}:
\begin{equation}\label{eq:TK}
    T_K \sim E_c \nep^{-\frac{\delta E(n_g) t_0}{2(M-2)\Gamma\Delta_P}}.
\end{equation}

The different behaviors at the charge degeneracy point ($n_g=0.5)$ and in the even Coulomb valley ($n_g=0.25$) are exemplified in Fig.~\ref{fig:Magnetization}(c), which displays $T_K$ versus $t_0/\Gamma$ for two values of $E_c$ and $M$ (see legend).
$T_K$ extracted at $n_g=0.5$ is independent of both $E_c$ and $M$ and it decreases with a power law compatible with $T_K\sim \Gamma$ (dashed line). 
For large values of $\Gamma$, the magnetization can change sign, preventing us from extracting $T_K$ with high precision (see the large errorbar at $n_g=0.5$ in Fig.~\ref{fig:Magnetization}(b)).
In the Coulomb valleys, instead, $T_K$ shows a substantial drop when increasing $E_c$: 
not only it is smaller for $E_c=0.4t_0$, but it decreases faster with $1/\Gamma$, in accordance with Eq. \eqref{eq:TK} (dot-dashed lines).
The data for $M=4$, $E_c=0.4t_0$ and $M=3$, $E_c=0.2t_0$ almost coincide as Eq.~\eqref{eq:TK} predicts the same behavior but for a factor 2 in front.
Our data display a concavity that is absent in Eq.~\eqref{eq:TK} and suggests a competing power law dependence on $\Gamma$ in agreement with NRG results of the low-energy effective model \cite{Galpin2014}. 
The presence of a power-law correction can also be understood as the interpolation between the scaling of $T_K$ deep in the Coulomb valleys, where it is dominated by the exponential decay, and that at $n_g=0.5$, where it is proportional to $\Gamma$. At intermediate values of the induced charge, $ 0 < n_g <0.5$, we expect a gradual crossover between the two regimes when the coupling strength $\Gamma$ becomes comparable with the charge excitations, as either $\Gamma$ increases or the charge degeneracy point is approached.

\paragraph{Nonlocal transport.-} To investigate multiterminal transport properties, we adopt a different quench protocol, using DMRG to prepare the ground state of the device coupled with $M$ leads  at equilibrium ($\mu_\alpha=0$). 
In general, such a state is a superposition of different charge and magnetization states.
At ${\sf t}=0$ we quench the chemical potential in the first lead to a finite value $\mu_1 = eV_b$ and compute the average current flowing through the remaining connected terminals. We refer to the latter as average nonlocal current.

\begin{figure}
    \centering
    \includegraphics[width=\columnwidth]{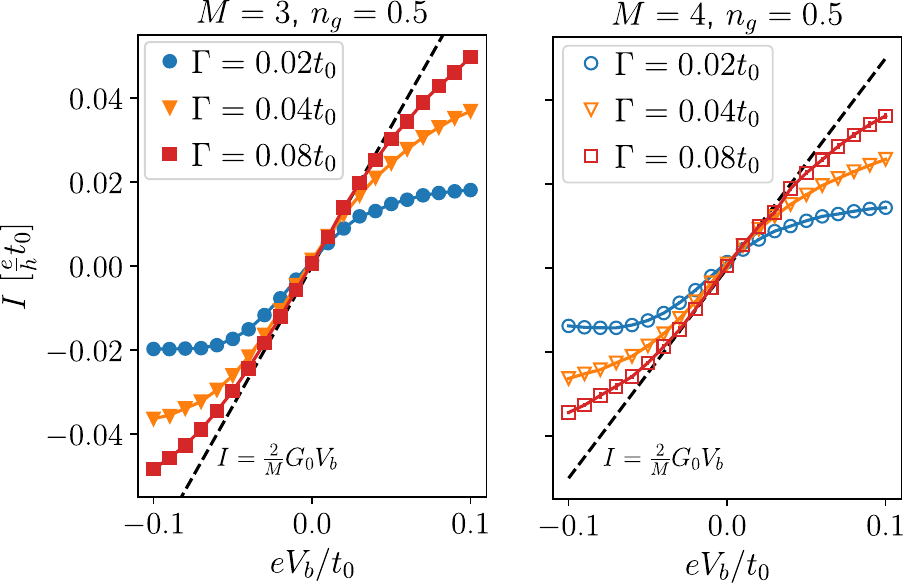}
    \caption{Average nonlocal current versus $V_b$ at $n_g=0.5$, for $M=3,\ 4$.
    The dashed line highlights the TKE prediction $G=\frac{2}{M}G_0$.
    The data are obtained with $\mathcal{L}=100$ and $\xi=32$.}
    \label{fig:current_Vb}
\end{figure}

RG predicts a fractional zero-bias nonlocal conductance, $G_{\rm TKE}=\frac{2}{M}\frac{e^2}{h}$, independent from all other physical parameters  for $T\ll T_K$, both in the Coulomb valleys\cite{Beri2012,Altland2013,Beri2013}, and at the charge-degeneracy points \cite{Mora2016,Michaeli2017,Papaj2019,Lutchyn2020}. 
Our simulations capture this fractional conductance for $M=3,4$ for sufficiently strong coupling in proximity of $n_g=0.5$, where $T_K$ is maximum and $G_{\rm TKE}$ can be observed for an extended voltage bias window (Fig. ~\ref{fig:current_Vb}). 
For $n_g \sim 0.5$, we also observe non-Fermi liquid power-law corrections with noninteger exponents which, however, do not seem compatible with the first-order scaling predicted by bosonization and RG~\cite{Zazunov2014,Mora2016,Michaeli2017,Beri2017, Papaj2019,Lutchyn2020,Supplemental}.

\begin{figure}
    \centering
    \includegraphics[width=0.8\columnwidth]{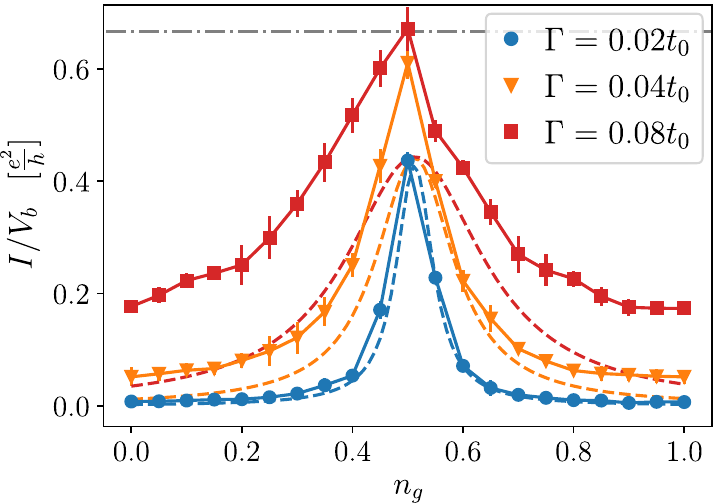}
    \caption{Nonlocal current versus $n_g$, for $E_c=0.4t_0$. The dashed lines are the corresponding RL approximation~\cite{Supplemental}. The horizontal dot-dashed line is $\frac{2e^2}{hM}$ with $M=3$.}
    \label{fig:current}
\end{figure}

Our simulations are performed at zero temperature, but, away from $n_g=0.5$, $T_K$ becomes comparable with the energy we introduce with the finite bias $eV_b$, competing with the universal strong-coupling features of the model.
In Fig.~\ref{fig:current} we plot the average nonlocal current ($M=3$ and $E_c=0.4t_0$) divided by the voltage bias as a function of $n_g$.
We set $\mu_1=eV_b=0.02t_0$, which is small enough to probe the response close to the linear regime, yet the data display a good signal-to-noise ratio.
The TKE prediction is met at the charge degeneracy point and strong coupling, consistently with Fig.~\ref{fig:current_Vb}, while the strong $n_g$ dependence confirms that we are not deep in the TKE regime; however, several hints of a strongly-correlated Kondo state emerge also in the Coulomb valleys.

In Fig. \ref{fig:current}, we compare our data with the conductance $G_{\rm RL}$ of a single fermionic resonant level (RL), which exhibits a peak scaling as $4G_0/M^2$ with width $\sim M\Gamma/E_c$ (dashed lines)~\cite{Supplemental}.
The data with the weakest coupling ($\Gamma=0.02t_0$) match well the RL approximation, as expected in a weak-coupling regime.
By increasing $\Gamma$, we observe large deviations from the single RL and the conductance rapidly approaches the TKE value of $\frac{2}{3}G_0$ (horizontal dot-dashed line) for  $n_g\sim0.5$. 
Indeed, in this regime, the applied voltage $\mu_1=0.02t_0$ is one order of magnitude smaller than the estimate of the Kondo temperature, $T_K \sim 0.1t_0$ in Fig.~\ref{fig:Magnetization}(c). 
Moreover, there is a substantial current flowing deep in the Coulomb valleys ($\Gamma=0.08t_0, 0.04t_0$) with apparent plateaus that suggest a crossover to the TKE regime. 
This is further confirmed by the analysis of the data averaged over the decay length $\xi$~\cite{Supplemental}. 

The main difficulty for our approach in investigating a deeper Kondo regime in the Coulomb valleys stems from the very low voltage bias needed to probe energy scales that are exponentially suppressed in $\delta E(n_g)$ as $T_K$. 
Indeed, finite-size effects limit the resolution in $eV_b$ that we can achieve and hinder the observation of Kondo physics at small coupling or deep in the Coulomb valleys.
Hence, our method is complementary to NRG: while the latter is more suited for studying the universal features of relevant low-energy approximations, our approach allows for a more direct comparison with experiments that have to cope with the interplay and possible competition between different energy scales.

In the supplemental materials~\cite{Supplemental}, we show the robustness of the TKE features against coupling anisotropies and the TKE disappearance when the MZMs acquire an energy splitting.

\paragraph{Conclusions.-} We analyzed the out-of-equilibrium properties of a minimal model for the topological Kondo effect, aiming at a microscopic description alternative to RG approaches and a qualitative understanding of transport signatures that may arise in double nanowire experiments.
Our results present evidence of the onset of strongly-correlated states compatible with a crossover between a weak-coupling and a topological Kondo regime.

First, the charge and the effective magnetization of the Majorana Cooper-pair box are characterized by different relaxation behaviors: the former only depends on the system-leads hybridization $\Gamma$, whereas the latter presents two separate timescales. 
In analogy with the dynamical features of the Anderson impurity model, we used the longer timescale to estimate the Kondo temperature associated with the TKE, with results compatible with the RG predictions \cite{Beri2012,Altland2013}. 

Second, the nonlocal conductance in the intermediate to strong-coupling regimes matches the predicted value $G_{\rm TKE} =2G_0/M$ at the charge degeneracy point, where $T_K$ is the largest. In the Coulomb valleys, it presents large deviations from the noninteracting resonant level approximation that well describes the weak-coupling regime and two-terminal devices~\cite{Chung_PRB2022}. 
When the resonant level approximation fails, the conductance displays a plateau in the Coulomb valleys, hinting at a crossover to the topological Kondo regime.

Our results are obtained through a MPS approach that allows for the study of topological Kondo models without recurring to perturbation theory in the Majorana-lead coupling nor requiring any particular hierarchy of the involved energy scales as expected in realistic devices \cite{Vekris_NanoLett2022}.
It is therefore well suited to understand the crossover between strong and weak-coupling regimes as well as the corrections to the RG predictions on the TKE when we probe the system at energy scales comparable with $T_K$.

Our method can be extended to more complex topological Kondo models, including the coupling of Majorana modes \cite{Altland2014} caused by crossed-Andreev and cotunneling processes, the generalization to multichannel systems \cite{Giuliano2013,Guanjie2023} and the presence of spurious quantum dots in the devices, which may cause additional Kondo effects \cite{Cheng2014,Lopez2014,Silva2020,Weymann2020,Svetogorov2023}. This approach can be applied to the transport features of many strongly interacting nanodevices based on quantum dots coupled to superconducting islands \cite{Liu2022,Tsintzis2022,Dvir2023,Tsintzis2023}.

\begin{acknowledgments}
\paragraph{Acknowledgements.-} We thank F. Buccheri, R. Egger and J. Paaske for insightful discussions.
M.W., L.M., and M.B. are supported by the Villum Foundation (Research Grant No. 25310). 
C.-M.C. acknowledges the support by the Ministry of Science and Technology (MOST) under Grant No. 111-2112-M-110-006-MY3, and by the Yushan Young Scholar Program under the Ministry of Education (MOE) in Taiwan.    
This project has received funding from the European Union’s Horizon 2020 research and innovation program under the Marie Skłodowska-Curie Grant Agreement No. 847523 “INTERACTIONS.”
This work was supported by Q@TN, the joint lab between University of Trento, FBK—Fondazione Bruno Kessler, INFN—National Institute for Nuclear Physics, and CNR—National Research Council.
\end{acknowledgments}

%

\newpage
\onecolumngrid
\section*{Supplemental Materials}

\section{Minimal two-sites Kitaev chain description}

In this Section we review the minimal double-nanowire description we use for modelling each nanowire and its comparison with standard constructions adopted to describe the topological Kondo effect.

The Majorana modes appearing in nanowire models that include spin-orbit and Zeeman coupling are characterized in general by a spinful wavefunction with position-dependent spinors \cite{Sticlet2012,Kjaergaard2012}. Therefore, a general tunnelling term combining these Majorana modes with spinful external leads couples them with a single related spin polarization; lead electrons in the orthogonal spin component, instead, are coupled by tunneling only to 
 high energy states beyond the gap of the superconducting system. This justifies the spinless approximation that we adopt in our model, since, in the low-energy sector, Majorana modes interact with a single fermionic species.


Differently from all previous investigations of the topological Kondo effect, we consider a minimal model that includes also Bogoliubov excitations and can be regarded as an effective zero-bandwidth description of the two superconducting nanowires. This minimal model corresponds to a two-site Kitaev chain for each nanowire.
It should be noted that this model can be extended for longer chains. The Hamiltonian for the chain $\sigma=\Up,\Dn$ is given by:
\begin{equation} \label{Akit}     
          \Ham^{(\sigma)}_{\rm kit}=\sum_{j=1}^{2}\left[ -\mu_{\rm sys}\,\opddag{j,\sigma}\opd{j,\sigma}+\left(-t_{\rm sys}\opddag{j+1,\sigma}\opd{j,\sigma}+  \Delta_P \nep^{i\Phi}\opd{j+1,\sigma}\opd{j,\sigma} +{\rm H.c.}\right)
      \right],
\end{equation}
where the index $j$ labels the site and $\Phi$ is the superconducting phase of the aluminum backbone. In our simulations, we set $t_{\rm sys}=\Delta_P=0.5t_0$ and $\mu_{\rm sys}=0.01t_0$ to avoid a perfect degeneracy of the states. The resulting energy splitting of the MZMs, however, is of order $10^{-4}t_0$ and is the smallest energy scale in our system. Before constructing the MPS representation of the full system, we diagonalize the quadratic Hamiltonians in Equation \eqref{Akit} and define the quasiparticle excitations.

For simplicity, in the following, we consider only the upper chain, as the two chains are indistinguishable at equilibrium. 

\begin{figure}[ht]
    \centering
    \includegraphics[width=0.5\columnwidth]{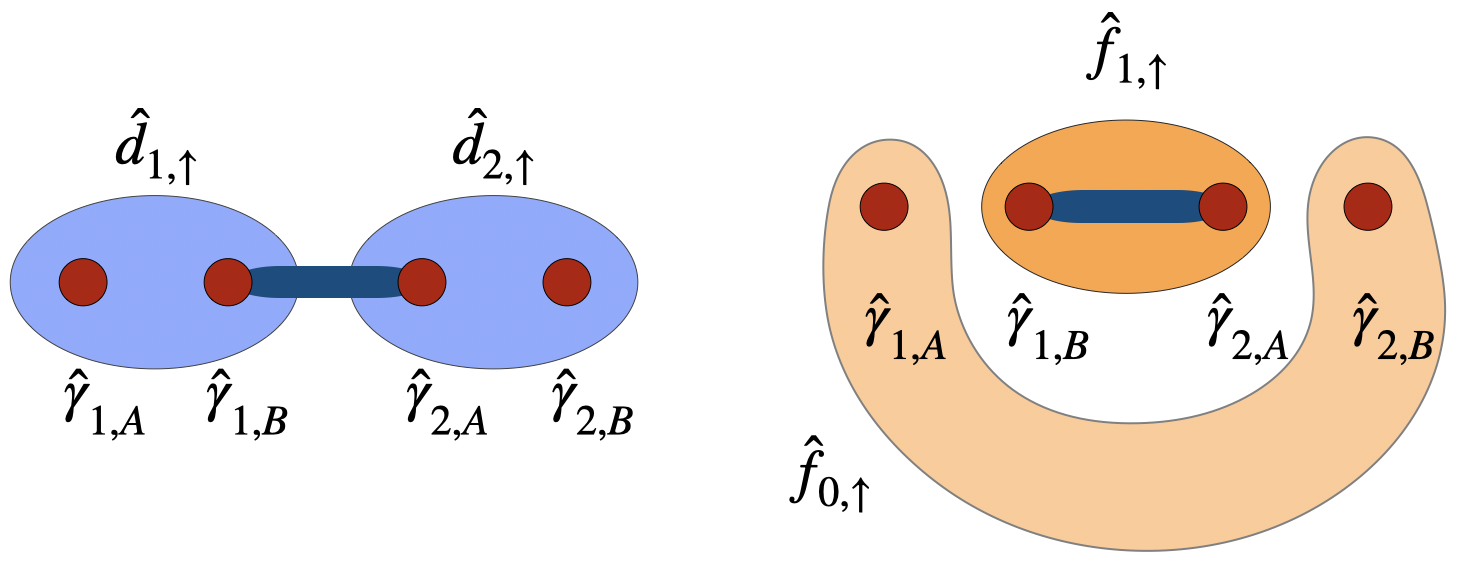}
    \caption{Each fermionic site of the Kitaev chain can be decomposed in two Majorana operators to make the MZMs physics more transparent. Quasiparticles states are represented schematically on the right. The thick blue link represents the interaction in Eq. \eqref{AHqp}.}
    \label{Afig:Kit}
\end{figure}

The physics of the MZMs is manifest by expressing each fermionic operator in terms of two Majorana fermions. In the case of the upper wire, we define:
\begin{equation} \label{Akitsubst}
    \opd{j,\uparrow}=\dfrac{\nep^{-i\Phi/2}}{2}\left(\opgamma{j,B}-i\opgamma{j,A}\right)
\end{equation}
as schematically represented in Fig.~\ref{Afig:Kit}. We consider the limit with $\mu_{\rm sys}=0$ and $\Delta_P=t_{\rm sys}$, such that the Hamiltonian is given by $\Ham^{(\uparrow)}_{\rm kit}=-i\Delta_P\opgamma{2,A}\opgamma{1,B}$ and couples Majorana fermions only at adjacent lattice sites (Fig.~\ref{Afig:Kit}). The ends of the chain support the unpaired MZMs $\opgamma{1}=\opgamma{1,A}$ and $\opgamma{2}=\opgamma{2,B}$ which allow us to define the zero-energy quasiparticle operator $\opf{0,\uparrow}$ introduced in the main text. In the quasiparticle basis, the first excited state has energy $\epsilon_{1,\uparrow}=2\Delta_P$ and corresponds to the operator $\opf{1,\uparrow}=(\opgamma{2,A} - i \opgamma{1,B})/2 $, such that the Hamiltonian can be written as 
\begin{equation}\label{AHqp}
    \Ham^{(\uparrow)}_{\rm Kit}= 2\Delta_P \left(\opfdag{1,\uparrow}\opf{1,\uparrow}-1/2\right).
\end{equation}
The same construction is valid for the lower chain.

To illustrate the relation of our construction with the standard models adopted for the description of the topological Kondo effect, we can consider the ideal topological limit with $\mu_{\rm sys}=0$, and restrict the model we analyzed to its low-energy sector.
The resulting effective coupling between the four Majorana zero-modes and the lead fermions is given by:
\begin{equation}
\Ham_{\rm t,0} = -\sum_{\alpha=1}^4 \frac{iJ_\alpha}{2} \hat{\gamma}_\alpha\left(\hat{c}_{\alpha,1}\nep^{i {\Phi}/2} + \hat{c}^\dag_{\alpha,1}\nep^{-i {\Phi}/2} \right)\,.
\end{equation}
This effective tunneling Hamiltonian corresponds to the coupling adopted in Refs. \cite{Altland2013,Beri2013} (with discretized lead degrees of freedom) to derive the topological Kondo effect. Away from the charge degeneracy point, by treating $\Ham_{\rm t,0}$ as a perturbation of the charging energy interaction in $\Ham_{\rm sys}$, one obtains the effective topological Kondo Hamiltonian term, quadratic in the Majorana modes, studied, for instance, in Ref. \cite{Beri2012}:
\begin{equation} \label{TKEeff}
\Ham_{\rm eff}= \sum_{\alpha\neq \beta } \lambda^+_{\alpha \beta} \hat{\gamma}_\alpha \hat{\gamma}_\beta \hat{c}^\dag_{\alpha,1} \hat{c}_{\beta,1} - \sum_\alpha \lambda_{\alpha \alpha}^- \hat{c}^\dag_{\alpha,1} \hat{c}_{\alpha,1}
\end{equation}
where $\lambda^{\pm}_{\alpha \beta} = \frac{J_\alpha J_\beta}{4} \left( \frac{1}{2E_C |n_g-1/2|} \pm \frac{1}{2E_C |n_g+1/2|} \right) $ for $0<n_g<1$ and $n_g$ sufficiently far from $1/2$.
The Majorana binomials in Eq. \eqref{TKEeff} define the effective Majorana magnetization in the three directions. Therefore, $\Ham_{\rm eff}$ can be recast in an anisotropic Kondo model. This mapping was adopted in Ref. \cite{Galpin2014} to study the TKE in the axial symmetric limit $J_1=J_2$ through NRG techniques.

\section{The matrix product state construction}\label{app:mps}

\subsection{Auxiliary charge site}

In order to simulate the topological Kondo model, we need to account for both its superconducting pairing and the charging energy of the Cooper-pair box. 
Importantly, the mean field BCS description of the superconducting system does not preserve the total particle number, but only its parity. This means that we cannot deduce the total charge of the Cooper-pair box $\hat{N}$ directly from the quasiparticle MPS construction. In order to overcome this problem we add an independent auxiliary charge site to the tensor network representation to keep track of the charge and its dynamics \cite{keselman2019,Chung_PRB2022}. 

First of all, one can promote the SC phase $\nep^{- i\Phi/2}$ in \eqref{Akitsubst} as the operator $\nep^{- i\hat{\Phi}/2}$ which lowers the number of electrons on the box by one (due to the charge phase relation $[\hat{N},\hat{\Phi}]=-2i$). In this way, the decomposition in Eq. \eqref{Akitsubst} enables to separate this charge degree of freedom from the quasiparticle number.  We can therefore describe charge dynamics by adding to our MPS an auxiliary site whose local Hilbert space is spanned by the eigenstates $\ket N$ of the charge $\hat{N}$ \cite{Chung_PRB2022}. 

The tunneling Hamiltonian $\Ham_{\rm t}$ becomes the sum of three-site operators of the form:
\begin{equation}
\label{AHt}
\Ham_{\rm t} = -\sum^4_{\alpha=1}\sum_{\sigma,n} {J_{\alpha}} \left[\nep^{i\hat{\Phi}/2} \left( u_{\alpha,\sigma,n} \opfdag{\sigma,n}+ v_{\alpha,\sigma,n} \opf{\sigma.n}\right)\opc{\alpha,1}+{\rm H.c.} \right].
\end{equation}
where the operator $\nep^{\pm i\hat{\Phi}/2}$ acts on the auxiliary site and raises/lowers the charge eigenvalue $N$.
Finally, charging energy costs are straightforwardly taken into account by considering the state of the auxiliary site, via $\Ham_{c}=E_c\left(\hat{N}-n_g\right)^2$ in Eq.~(1) in the main text. 

The auxiliary charge site construction is numerically implemented by restricting its local Hilbert, $N\in~\left[-N_{\rm max}, N_{\rm max}\right]$, with $N_{\rm max}=5$ (such that $\nep^{\pm i \hat{\Phi}/2}$ are represented as $11\times 11$ matrices). Moreover, to remove the redundancy introduced by the auxiliary site, we constrain the parity of $\hat{N}$ to be the same as the parity of the total occupation of the quasiparticle states in the Majorana Cooper-pair box \cite{Chung_PRB2022}. Namely, once defining the operator
\begin{equation}
    \widehat{P}=\left(-1\right)^{\hat{N}+\sum_{n,\sigma}\opfdag{n,\sigma}\opf{n,\sigma}},
\end{equation}
the following relation 
\begin{equation}
    \widehat{P}\ket{\psi_{\rm phys}}=\ket{\psi_{\rm phys}},
\end{equation}
has to be valid for any physical states $\ket{\psi_{\rm phys}}$.
Our MPS and matrix product operator construction encodes such $\mathbb{Z}_2$ constraint.

\subsection{TDVP Dynamics and transport quantities}

We simulate the dynamics of the system through the TDVP algorithm, which is not limited by the long-range Hamiltonian resulting from the energy basis choice for the MPS. The Hamiltonian $\Ham$ is represented as a matrix product operator of maximum bond dimension $\chi=16$.

\begin{figure}
    \centering
    \includegraphics[width=0.45\columnwidth]{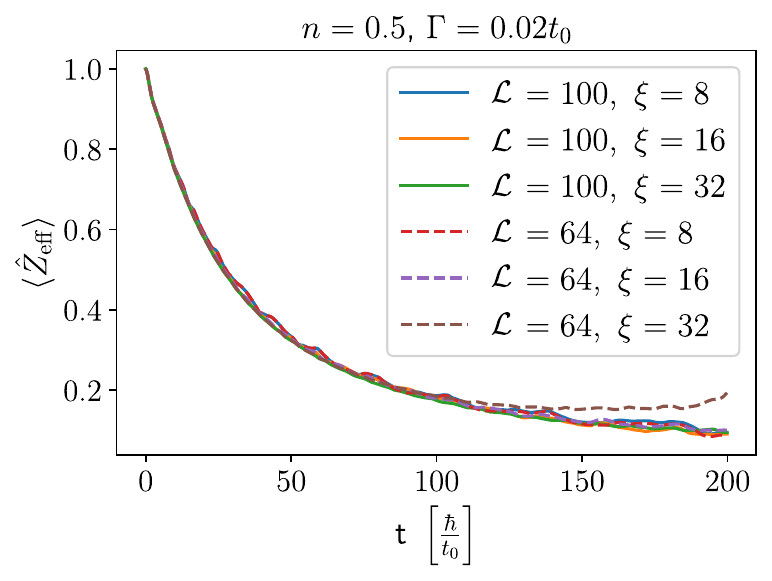}
    \caption{Relaxation of the effective magnetization at the charge degeneracy point for different combinations of lead length $\mathcal{L}$ and hopping decay length $\xi$.}
    \label{fig:convergence}
\end{figure}

A crucial element of our simulation is the exponential decay of the hopping amplitude inside the leads. It allows for both a tunable energy resolution resolution at small bias and for simulating an effectively longer system by slowing down the current as it propagates in the leads~\cite{Chung_PRB2022}. 
To ensure that we get physical results, a first step in our simulations is to verify the convergence of the dynamics with respect to the lead length $\mathcal{L}$ and the hopping decay length $\xi$.
We report in Fig.~\ref{fig:convergence} an example of such calculation, where we plot the effective magnetization as function of time for different choices of $\mathcal{L}$ and $\xi$.
Beside the combination $\xi=32,\, \mathcal{L}=64$, all other curves collapse perfectly on each other, showing that the simulations are converged to a physically meaningful dynamics.
The deviation from the physical behaviour is expected to take place at ${\sf t}_{\rm max}\simeq \frac{\hbar}{t_0} \frac{\xi}{2}\left( \nep^{\mathcal{L}/\xi}-1\right)$~\cite{Chung_PRB2022}, which is approximately $100{\hbar}/t_0$ for $\mathcal{L}=64,\, \xi=32$, as confirmed by the simulations.

\begin{figure}[ht]
    \centering
    \includegraphics[width=0.6\columnwidth]{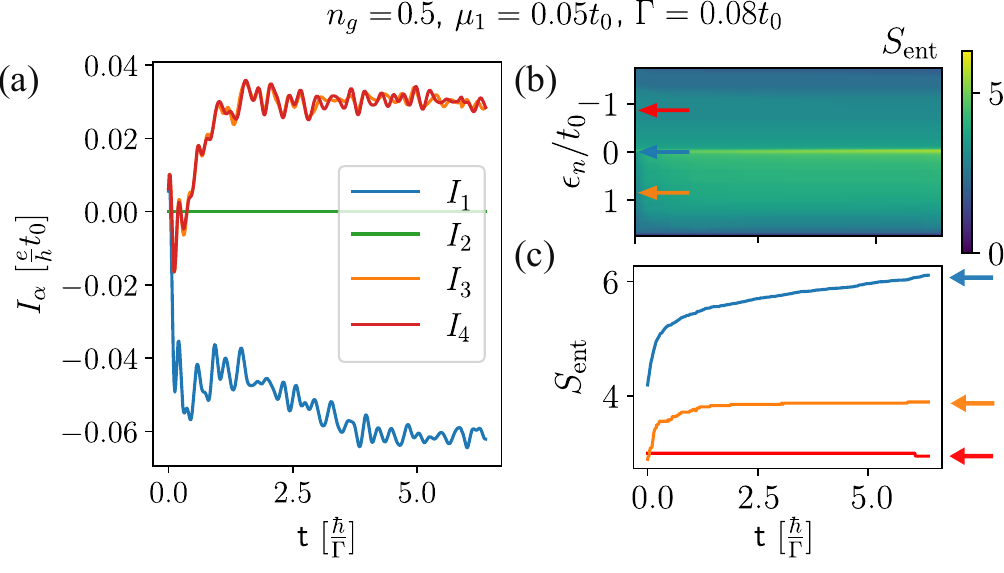}
    \caption{Typical dynamics of the current and entanglement entropy after a quantum quench for a three-terminal device.
    (a) Time dependence of the current on each lead. $I_1$ has a negative sign because it is the only in-going current. $I_2=0$ because the corresponding lead is decoupled from the device.
    (b) Entanglement entropy at each bond of the MPS as a function of energy and time. The three arrows mark the horizontal line cuts corresponding to the curves shown in panel (c).
    Simulation parameters: $\mathcal{L}=100$, $\xi=32$.} 
    \label{Afig:dynamics_example}
\end{figure}

Once convergence is ensured, $\mathcal{L}$ and $\xi$ are chosen in such a way that the current signal displays a long enough pleateau for ${\sf t} < {\sf t}_{\rm max}$ . When these conditions are met, we can characterize the emerging  non-equilibrium quasi-steady states, that provide faithful descriptions of the physical behavior of the (infinite) system in its stationary state (see, for instance, Refs. \cite{bertini2016,Essler_2016}).
An example of the current after the quench is depicted in Fig. \ref{Afig:dynamics_example}(a). To extract the values of the currents analyzed in the main text, we average the signal after it reaches the stationary value and estimate the errorbars through standard binning techniques. 

Thanks to the chosen basis, after the quantum quench, the entanglement entropy of the system increases logarithmically with time \cite{Rams2020,Chung_PRB2022} and it is mainly localized in an energy window proportional to the voltage bias. Outside that window, the entanglement entropy is mostly time-independent, as we show in Fig.~\ref{Afig:dynamics_example}(b) and (c).
See Ref.~\cite{Chung_PRB2022} for more details about the MPS construction and dynamics.

We finally observe that the Hamiltonian we adopt to describe the double-nanowire model displays an additional symmetry with respect to the most common topological Kondo models \cite{Beri2012,Altland2013,Beri2013}. Indeed, the dynamics we analyze separately preserves the two fermionic parities:
\begin{align}
&\hat{P}_\Up=(-1)^{\sum_{\alpha=1,2}\sum_{l=1}^{\mathcal{L}}\hat{c}^\dag_{\alpha,l}\hat{c}_{\alpha,l}+\sum_{n=0,1}\opfdag{n,\Up}\opf{n\,\Up}}\,,\\
&\hat{P}_\Dn=(-1)^{\sum_{\alpha=3,4}\sum_{l=1}^{\mathcal{L}}\hat{c}^\dag_{\alpha,l}\hat{c}_{\alpha,l}+\sum_{n=0,1}\opfdag{n,\Dn}\opf{n\,\Dn}}\,.
\end{align}
These symmetries reflect the fact that we are neglecting crossed-Andreev and direct cotunneling processes mediated by the superconducting island between the two nanowires. These conservations have the important effect of breaking the particle-hole-like symmetry of the dynamics between systems characterized by $n_g$ and $1-n_g$, as can be seen from Fig.~5 in the main text at large coupling. The initial ground states $\ket{00}$ ($N=0$) for $n_g <0.5$ and $\ket{10}$ ($N=1$) for $n_g>0.5$ correspond to different sectors of $\hat{P}_\Dn$ and are not mapped one into the other by the symmetry.

\section{Further transport results}
\subsection{Asymmetric couplings}
Here we investigate the effect of introducing an asymmetry in the couplings ${\Gamma_\alpha}$ on the transport properties.

\begin{figure}
    \centering
    \includegraphics[width=0.45\columnwidth]{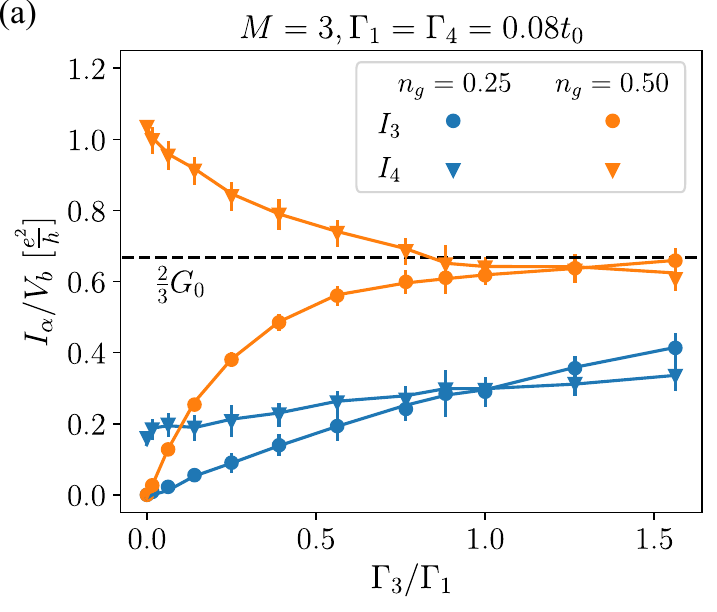}
    \includegraphics[width=0.45\columnwidth]{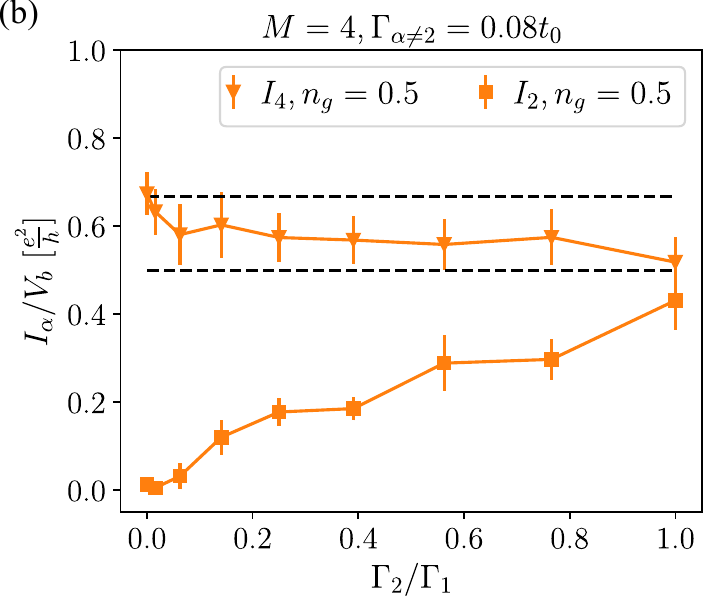}
    \caption{(a) Nonlocal current as a function of the varying coupling strength $\Gamma_3$, at the charge degeneracy point $n_g=0.5$ and in the even valley $n_g=0.25$. Lead 2 is uncoupled ($\Gamma_2=0$).
   (b) Nonlocal currents as a function of the varying coupling strength $\Gamma_2$, at the charge degeneracy point $n_g=0.5$. The two horizontal dashed lines mark $G=\frac{2}{M} G_0$ for $M=3,4$.
   In both panels $E_c=0.4$, $eV_b =0.01t_0$, $L=100$, and $\xi=32$.}
    \label{fig:curren_M3_asym}
\end{figure}

We first analyze a device with three leads ($\Gamma_2=0$), where two of them have the same coupling strength $\Gamma_1=\Gamma_4=0.08t_0$, which corresponds to the strong coupling regime explored in the main text, while the third is varied. 
In Fig.~\ref{fig:curren_M3_asym}(a) we plot the nonlocal currents $I_3$ and $I_4$ divided by the bias on lead 1 as we vary $\Gamma_3 \in [0,1.6 \Gamma_1]$.
At the charge degeneracy point (orange symbols), the data suggest that the current is approximately stable for a broad range of couplings $\Gamma_3 \gtrsim \Gamma_1$, and, within the error bars, is compatible with the linear conductance associated to the TKE (horizontal dashed line).
As $\Gamma_3$ decreases, $I_3$ also decreases and vanishes when the lead is finally decoupled from the system. At the same time, $I_4$ increases and approaches the quantized value $I_4=G_0 V_b$ when the device has only two terminals, as we expect from the resonant tunneling mediated by MZMs with symmetric couplings~\cite{vanHeck_PRB2016,Chung_PRB2022}.
As we move deeper in the Coulomb valley ($n_g=0.25$, blue symbols), the system appears to be further away from the TKE regime and the current shows a roughly linear dependence on $\Gamma_3$. 
Interestingly, however, the current $I_4$ decreases upon switching off $\Gamma_3$, despite keeping $\Gamma_4$ constant. This is in contrast with the single resonant level prediction of Eq.~\eqref{eq:G}, confirming that a contribution to the current originating from a strongly coupled state is present also in the Coulomb valleys, even though the TKE quantization of the conductance is not recovered for the chosen parameter ranges.


Let us now focus on the crossover between $M=4$ and $M=3$: we consider a four-terminal device where we tune the coupling $\Gamma_2$ from the symmetric configuration, $\Gamma_\alpha=0.08t_0$ on any lead, to $\Gamma_2=0$, while keeping a small voltage bias $eV_b=0.01 t_0$ on lead 1. 
In Fig.~\ref{fig:curren_M3_asym}(b) we show the nonlocal currents $I_2$ and $I_4$ for $n_g=0.5$, where the Kondo temperature is maximal, as we switch off $\Gamma_2$.
When $\Gamma_2=\Gamma_1$, the current on both leads is again compatible with the TKE prediction with $M=4$ (lower horizontal dashed line). 
As $\Gamma_2$ decreases, $I_2$ and $I_4$ display opposite behaviors; the former decreases and vanishes following $\Gamma_2$ while the latter displays first a rather flat plateau followed by a rapid increase to match the TKE prediction for $M=3$ when $\Gamma_2 \to 0$ (higher horizontal dashed line).
The current on lead 3 (data not shown) follows closely the signal on $I_4$.

\subsection{Low bias transport}\label{app:lowbias}
\begin{figure}
    \centering
    \includegraphics[width=0.45\columnwidth]{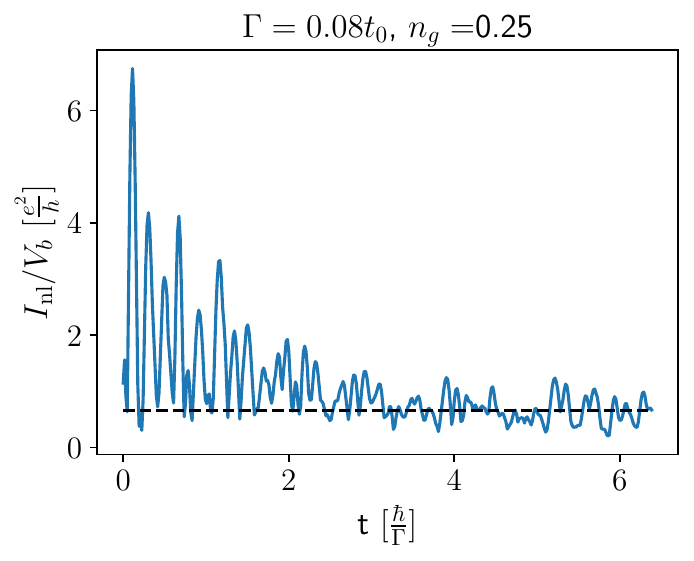}
    \caption{Time dependence of the nonlocal current for $M=3$ and $n_g=0.25$, averaged over 4 values of $\xi=2,4,8,16$. The chosen bias is $eV_b=0.001t_0$. The horizontal dashed line marks the TKE prediction. }
    \label{fig:current_ng025}
\end{figure}

Transport simulations at very low bias are hampered by a low signal-to-noise ratio that prevents from an accurate estimate of the average current in the nonequilibrium quasi-steady state.
This limitation is relevant for low Kondo temperatures as, for instance, in the Coulomb valleys. 
To partially circumvent this issue, inspired by the so-called z-trick \cite{Yoshida1990} commonly used in NRG methods, here we consider data obtained by averaging over different logarithmic discretizations of the energy levels of the leads. In particular, we average the currents over different decay lengths of the hopping amplitude in the leads.
In Fig.~\ref{fig:current_ng025} we plot an example of this procedure: we consider $n_g=0.25$ (even-parity Coulomb valley), $E_c=0.4t_0$, $M=3$, and $\Gamma=0.08t_0$. 
The corresponding Kondo temperature extracted from the magnetization dynamics is $T_K\sim 0.01t_0$, see Fig.~3(c) of the main text.
To capture the transport signature of the TKE, we perform different simulations with a small bias $eV_b=10^{-3}t_0$ on lead 1 and $\xi=\lbrace 2,4,8,16\rbrace$. 
We then average the outgoing current over the different values of $\xi$.
This reduces the amplitude of the current oscillations, and leads to a good match with the TKE prediction $G_{\rm TKE}=\frac{2}{M}G_0$ also in the Coulomb valleys.

In the main text, we compare the extracted low bias conductance with the resonant level approximation. This consists in one single charge degree of freedom $\hat{n}$ affected by a charging energy $E_c\left(\hat{n}-n_g\right)^2$ and coupled with $M$ leads with symmetric tunneling rates $\Gamma_\alpha=\Gamma$. For a finite bias difference $\mu$ between leads, the non-local conductance is given by
\begin{equation}\label{eq:G}
    G_{\rm RL}(n_g,\mu) = \frac{e^2}{h}\frac{4\Gamma^2}{M^2\Gamma^2 + 4[\mu-E_c(1-2n_g)]^2} \ .
\end{equation}
This approximation gives the corresponding dashed lines plotted in Fig.~5 of the main text, for different values of the tunneling rate.


\subsection{Finite bias corrections}\label{app:biascorr}

\begin{figure}
    \centering
    \includegraphics[width=0.45\columnwidth]{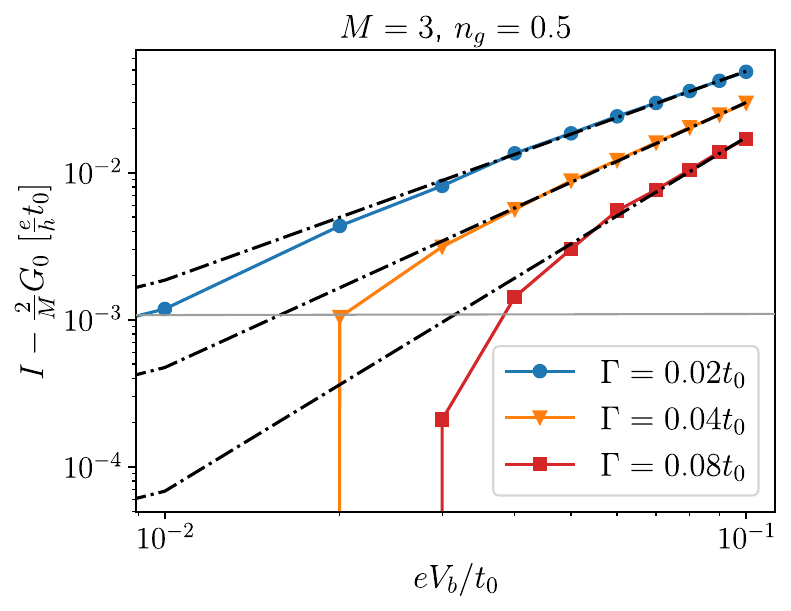}
    \includegraphics[width=0.45\columnwidth]{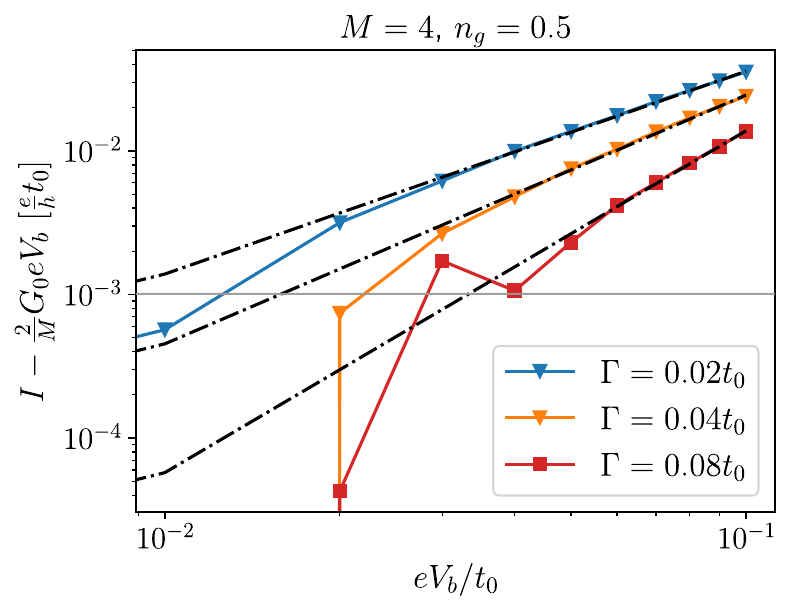}
    \caption{Deviation of the average non-local current from the linear regime $I=\frac{2}{M}G_0 V_b$, for $M=3,4$, $n_g=0.5$ and $E_c=0.4t_0$. The black dashed lines are power-law fits $bx^a$ of the data. The values of $a$ for $M=3$ are 1.42(1) ($\Gamma=0.02t_0$), 1.80(1) ($\Gamma=0.04t_0$), and 2.41(3) ($\Gamma=0.08t_0$), while for $M=4$ we obtain 1.41(2) ($\Gamma=0.02t_0$), 1.73(2) ($\Gamma=0.04t_0$), and 2.38(4) ($\Gamma=0.08t_0$). The grey line indicates the numerical precision.}
    \label{fig:biascorr}
\end{figure}

Finally, we discuss the finite bias corrections to the currents close to TKE linear response behaviour, $I=\frac{2}{M}G_0 V_b$. In a renormalization group sense, a power law correction $G= 2G_0/M - AV_b^\alpha$ is related to the scaling dimension of the most relevant operator which arises at the TKE fixed point. In particular, a fixed point described by the Fermi liquid (FL) theory displays a quadratic correction for the conductance ($\alpha=2$). The topological Kondo effect, instead, is predicted to display non-Fermi liquid corrections defined by the universal fractional exponent $\alpha=2(1-2/M)$\cite{Mora2016,Zazunov2014,Beri2017}.
The boundary CFT approach \cite{Affleck1990,Affleck1991} (adapted to the $SO(M)$ case) can also provide a full description of the scaling of these perturbations close to the strong-coupling fixed point. Additionally, techniques based on the mapping on boundary sine-Gordon models have been adopted to study the universal features of the transport at finite bias and temperature \cite{Beri2017}.

In Fig.~\ref{fig:biascorr}, we show the bias dependence of the current deviation from the TKE regime, $I-\frac{2}{M}G_0 V_b$, for $M=3,4$. The data show a clear power-law behavior, particularly for bias values $V_b$ that are not excessively small (such that the signal-to-noise ratio is reliable). In the displayed cases, the deviation of the current from the power-law fits is below the numerical precision $\sim10^{-3}$.

In all cases, we observe a non-Fermi liquid scaling which significantly deviates from the cubic FL behavior of the current $V_b^3$. However, the fitted exponents do not match the RG predicted values $\alpha+1 = 3-4/M$. The absence of a clear separation of energy scales in the problem might be a source of deviation from the perturbative RG analysis. Moreover, close to the charge degeneracy point, intermediate fixed points are believed to emerge \cite{Mora2016}. Finally, the fitted exponents seem to depend continuously on the coupling strength $\Gamma$ and, while this analysis has shed light on a non-Fermi liquid behavior, further analysis is needed to understand these power-law corrections. 

\subsection{Breakdown of the TKE caused by a Majorana energy splitting}
Here we investigate the effect on low-bias transport of an energy splitting between the subgap states of the lower and upper nanowires.
In our modeling, this splitting is achieved by increasing the chemical potential $\mu_{\rm sys}$ of the Kitaev chain of the lower wire, coupling the MZMs together and raising the energy of the corresponding fermionic level to $\epsilon_{0,\downarrow}=0.08t_0$. The energy  of the subgap state of the upper wire is $\epsilon_{0,\uparrow}\sim10^{-5}t_0$, as in all other numerical results.
We consider $M=3$ leads and two different coupling configurations: in configuration A, both MZMs in the upper wire are coupled to the corresponding leads ($\alpha=1,\ 2$), while the trivial subgap state in the lower wire is coupled to a single grounded lead; in configuration B, only the biased lead ($\alpha=1$) is coupled with the upper wire, while the lower one has both leads 3 and 4 attached.
\begin{figure}
    \centering
    \includegraphics[width=0.45\columnwidth]{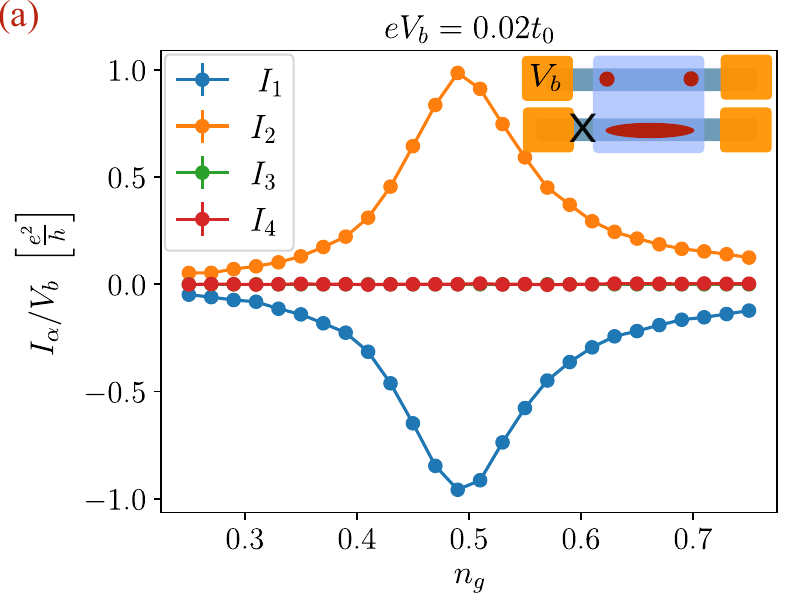}
    \includegraphics[width=0.45\columnwidth]{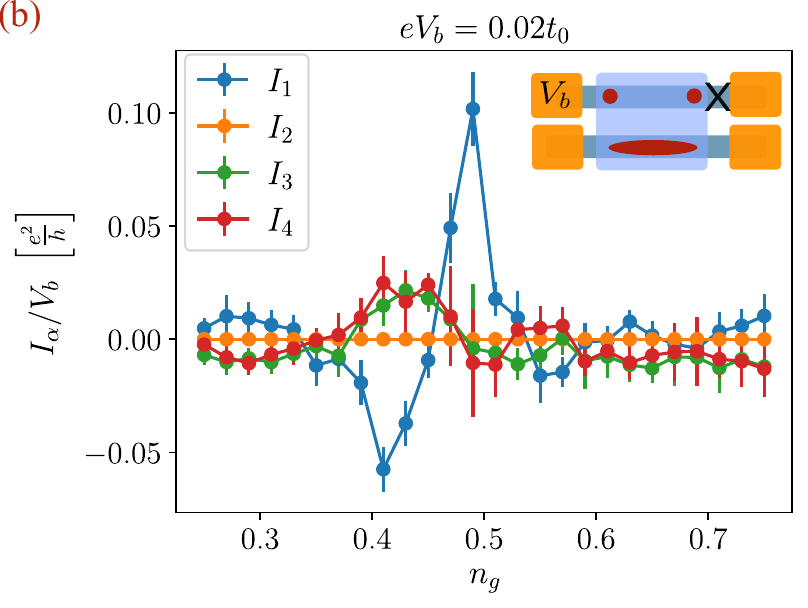}
    \caption{Current as a function of the induce charge $n_g$ around the charge degeneracy point. A small voltage bias $eV_b=0.02t_0$ is set on lead 1. In the upper nanowire the lowest energy state is a superposition of two MZM while in the lower one there is a subgap state of energy $\epsilon_{0,\downarrow}=0.08t_0$. The coupling with the connected leads is $\Gamma=0.04t_0$
    (a) configuration A, with lead 3 decoupled. (b) configuration B, with lead 2 decoupled. Notice the different amplitude of the current in the two panels. }
    \label{fig:TKE_breakdown}
\end{figure}

The results for configuration A are shown in Fig.~\ref{fig:TKE_breakdown}(a), where we plot the current on all four leads as a function of the induced charge $n_g$. The voltage bias $eV_b=0.02t_0$ is smaller than the energy $\epsilon_{0,\downarrow}$ of the subgap state in the lower wire.
Consequently, there is almost no current flow in lead 4 (the only coupled to the lower wire) and the system behaves effectively as a two terminal device. Indeed, $I_{1/2}/V_b \simeq e^2/h$ at the charge degeneracy point, as we expect from resonant tunneling through a single pair of MZMs \cite{vanHeck_PRB2016}. 
The asymmetry of the peak around $n_g=0.5$ is due to faster cotunneling processes in the odd-parity Coulomb diamond (see Ref. \cite{Chung_PRB2022}). $I_1$ and $I_2$ are almost identical but for the sign, since $I_1$ is an ingoing current, while $I_2$ is outgoing.

Figure \ref{fig:TKE_breakdown}(b) shows our results for configuration B. First, observe that the amplitude of the current signal is suppressed by a factor of 10 with respect to configuration A, missing a resonant state that can mediate transport. 
Moreover, the current sign changes multiple times as we tune $n_g$, with some similarities to Ref.~\cite{Souto_PRB2022}, as different processes become dominant when the energies of the SC island many-body states cross the Fermi level of the leads. 
Furthermore, the equilibration time also increases, with respect to configuration A, maybe due to slow charge fluctuations between the leads and the superconductor.
In both cases, the current signal we observe is markedly different from the corresponding behavior for the TKE, Fig.~[5] of the main text, as the degeneracy at the heart of TKE is broken.
In an experimental setup, an accidental degeneracy between trivial subgap states would be easy to remove by tuning external gates, while a robust observation of the features presented in the main text would be a strong smoking gun for the presence of MZMs in the system.

\begin{figure}
    \centering
    \includegraphics[width=0.5\columnwidth]{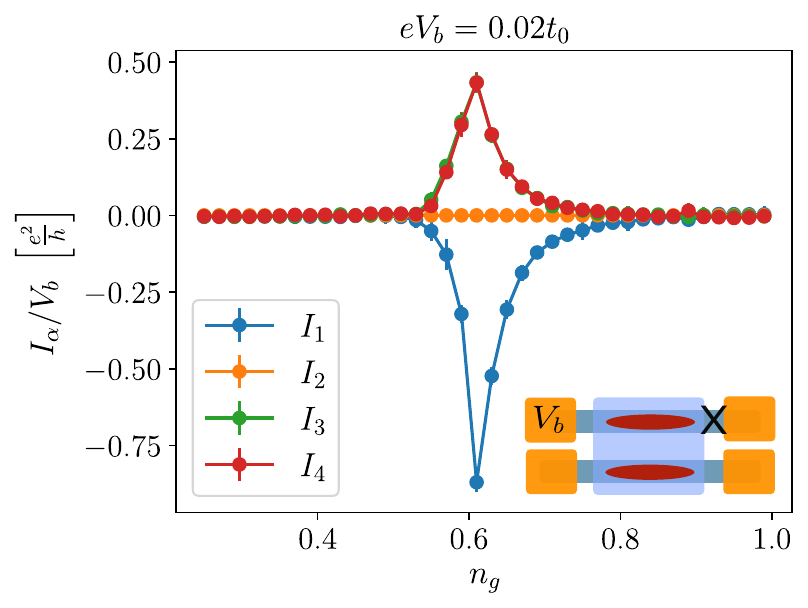}
    \caption{Local and nonlocal linear conductance $I_\alpha/\mu_1$ as a function of $n_g$, for a configuration with $M=3$ leads and subgap states with a finite energy $\epsilon_{0,\sigma}=0.08t_0$.}
    \label{fig:current_ABS}
\end{figure}

Finally, we investigate transport in a configuration where the upper and lower nanowires are degenerate but with a finite overlap between the MZMs, thus creating finite-energy subgap states in both wires.
In particular, we consider $\epsilon_{0,\sigma}=0.08t_0$ for both wires and $M=3$ leads, where $\Gamma_2=0$ and $\Gamma_{\alpha\neq 2}=0.04t_0$.
Figure \ref{fig:current_ABS} reports the current on each lead as a function of $n_g$. 
With respect to the corresponding data in Fig.[5] of the main text, there is a clear shift of the charge degeneracy point to $n_g \sim 0.6$ due to the finite energy of the state mediating transport.
However, the most striking difference is the strong suppression of the current far from the charge resonance. On the contrary, when the current is mediated by MZM and the TKE is expected to take place, a sizable nonzero current flows also deep in the Coulomb valley, even when transport is probed at energies above the $T_K$.

\end{document}